\documentclass[conference]{IEEEtran}
\IEEEoverridecommandlockouts
\usepackage{cite}
\usepackage{tikz}
\usepackage{notoccite}
\usepackage{amsmath,amssymb,amsfonts}
\usepackage{algorithmic}
\usepackage{graphicx}
\usepackage{textcomp}
\usepackage{xcolor}
\usepackage[top=72pt,left=54pt,right=54pt,bottom=54pt,includehead,includefoot]{geometry}

\usepackage{afterpage}

\makeatletter
    \newcommand{\linebreakand}{%
      \end{@IEEEauthorhalign}
      \hfill\mbox{}\par
      \mbox{}\hfill\begin{@IEEEauthorhalign}
    }
    \makeatother   
\newcommand\copyrighttext{%
  \footnotesize \textcopyright 2023 IEEE. Personal use of this material is permitted. Permission from IEEE must be obtained for all other uses, in any current or future media, including reprinting/republishing this material for advertising or promotional purposes, creating new collective works, for resale or redistribution to servers or lists, or reuse of any copyrighted component of this work in other works.
  DOI: Will be updated soon}
\newcommand\copyrightnotice{%
\begin{tikzpicture}[remember picture,overlay]
\node[anchor=south,yshift=10pt] at (current page.south) {\fbox{\parbox{\dimexpr\textwidth-\fboxsep-\fboxrule\relax}{\copyrighttext}}};
\end{tikzpicture}%
}

\begin{document}

\title{Emotion Detection from EEG \\using Transfer Learning
{

}}

% \name{Sidharth Sidharth\IEEEauthorrefmark{1}$^{1}$, Ashish Abraham Samuel\IEEEauthorrefmark{1}$^{1}$, Ranjana H\IEEEauthorrefmark{1}$^{1}$, Jerrin Thomas Panachakel$^{1}$, Sana Parveen K$^{1}$}

\author{
  \IEEEauthorblockN{Sidharth Sidharth\IEEEauthorrefmark{1}\IEEEauthorrefmark{2},}
\IEEEcompsocitemizethanks{\IEEEcompsocthanksitem\IEEEauthorrefmark{1}Sidharth Sidharth, Ashish Abraham Samuel and Ranjana H are co-first authors}
\IEEEcompsocitemizethanks{\IEEEcompsocthanksitem\IEEEauthorrefmark{2}Department of Electronics and Communication Engineering, College of Engineering Trivandrum, India.(tve19ae051@cet.ac.in)}
  % \IEEEauthorblockA{\textit{Dept. of Electronics and Comm. Engg.} \\
  %   \textit{College of Engineering}\\
  %   Trivandrum, India \\
  %   tve19ae051@cet.ac.in}
  \and
  \IEEEauthorblockN{Ashish Abraham Samuel\IEEEauthorrefmark{1}\IEEEauthorrefmark{2},}
  % \IEEEauthorblockA{\textit{Dept. of Electronics and Comm. Engg.} \\
  %   \textit{College of Engineering}\\
  %   Trivandrum, India \\
  %   tve19ae016@cet.ac.in}
  \and
  \IEEEauthorblockN{Ranjana H\IEEEauthorrefmark{1}\IEEEauthorrefmark{2},}
  % \IEEEauthorblockA{\textit{Dept. of Electronics and Comm. Engg.} \\
  %   \textit{College of Engineering}\\
  %   Trivandrum, India \\
  %   tve19ae043@cet.ac.in}   \linebreakand % <------------- \and with a line-break
  \and
   \IEEEauthorblockN{Jerrin Thomas Panachakel\IEEEauthorrefmark{2},}
  % \IEEEauthorblockA{\textit{Dept. of Electronics and Comm. Engineering} \\
  %   \textit{College of Engineering}\\
  %   Trivandrum, India \\
  %   jerrin.panachakel@cet.ac.in} 
  \and
\IEEEauthorblockN{Sana Parveen K\IEEEauthorrefmark{2},}
  % \IEEEauthorblockA{\textit{Dept. of Electronics and Comm. Engg.} \\
  %   \textit{College of Engineering}\\
  %   Trivandrum, India \\
  %   tve21ecsp13@cet.ac.in}

}

\maketitle
\copyrightnotice
\begin{abstract}
The detection of emotions using an Electroencephalogram (EEG) is a crucial area in brain-computer interfaces and has valuable applications in fields such as rehabilitation and medicine. In this study, we employed transfer learning to overcome the challenge of limited data availability in EEG-based emotion detection. The base model used in this study was Resnet50. Additionally, we employed a novel feature combination in EEG-based emotion detection. The input to the model was in the form of an image matrix, which comprised Mean Phase Coherence (MPC) and Magnitude Squared Coherence (MSC) in the upper-triangular and lower-triangular matrices, respectively. We further improved the technique by incorporating features obtained from the Differential Entropy (DE) into the diagonal, which previously held little to no useful information for classifying emotions. The dataset used in this study, SEED EEG (62 channel EEG), comprises three classes (Positive, Neutral, and Negative). We calculated both subject-independent and subject-dependent accuracy. The subject-dependent accuracy was obtained using a 10-fold cross-validation method and was 93.1\%, while the subject-independent classification was performed by employing the leave-one-subject-out (LOSO) strategy. The accuracy obtained in subject-independent classification was 71.6\%. Both of these accuracies are at least twice better than the chance accuracy of classifying 3 classes. The study found the use of MSC and MPC in EEG-based emotion detection promising for emotion classification. The future scope of this work includes the use of data augmentation techniques, enhanced classifiers, and better features for emotion classification.
\end{abstract}


\section{Introduction}
Emotion detection from biological signals has become a highly significant area of study in the realms of neuroscience and signal processing research \cite{parveen2023}. Emotion detection could be done in various ways. Such as through facial expressions, paralinguistic analysis, physiological signals, behavioral signals, and biochemical signals.

Electroencephelography (EEG) is a method of measuring the electrical activity of the brain and is widely used for emotion detection. It possesses numerous advantages over other methods. Being a non-invasive method, it does not deliver an unpleasant experience to the user. It can provide a high temporal and spatial resolution (subjective to the number of electrodes used) \cite{Burle2015_EEG_res} of brain activity that enables the detection of rapid changes in an emotional state in a specific section of the brain. It is easily portable. Moreover, EEG is a direct measurement of brain activity, which provides an objective measure of the emotional state. \\
The study \cite{Lee} demonstrated the utility of coherence as a connectivity index between EEG channels in understanding the relationship between brain activity and emotional states.
Previous research in the field of emotion detection from EEG signals has employed various techniques for feature extraction, including the Short Time Fourier Transform (STFT)\cite{Iyer}, Differential Entropy (DE)\cite{Wang}, Power Spectral Density (PSD)\cite{Demir}, and Continuous Wavelet Transform (CWT)\cite{Review}.\\
Emotion is a multi-faceted and intricate process that engages various regions of the brain. As such, utilizing simultaneous feature extraction from multichannel EEG data is crucial to identify an emotion accurately. Few works which include \cite{GAO2022105606} have already employed simultaneous feature extraction from EEG for emotion detection. \cite{GAO2022105606} employed the Riemannian geometry properties of the Symmetric Positive Definite (SPD) matrix obtained from the original EEG data to classify emotions. \\
In \cite{panachakel}, mean phase coherence (MPC) and magnitude squared coherence (MSC) was used as features for decoding imagined speech and also employed transfer learning along with data augmentation to overcome the data scarcity issue.\\
In our work, we used MPC and MSC as the main features employed in emotion classification. The dataset used was SEED\_EEG\cite{7104132}. It classifies emotions into three categories; positive, negative, and neutral. We also leveraged the information extracted from Differential Entropy (DE) provided in the dataset. We adopted ResNet-50 based transfer learning model as the classifier to tackle the problem of data insufficiency. The concept of using MPC and MSC as features in classifying emotions using transfer learning is the novel method presented in this work.
\section{Dataset}
The dataset used for this study is SEED (SJTU Emotion EEG Dataset). We specifically utilized the SEED\_EEG data which contains the EEG of 15 subjects(7 males and 8 females). They were shown approximately 4 minutes of movie clips that were chosen and edited to induce positive, negative, or neutral emotions. Each subject underwent the experiment 3 times within an interval of one week, giving a total of 45 trials.
The data were downsampled to 200 Hz following which a bandpass filter of 0-75 Hz was applied. 62-channel EEG was obtained via the international 10 - 20 system EEG cap using brain-like computing and machine learning(BCMI) methods. 
The 62-channel EEG data is converted into three frequency bands - alpha (8 to 13 Hz), beta(13 to 30 Hz), and gamma(30 to 70Hz) by using a bandpass filter. MPC and MSC features are extracted for each band and entered into an array such that the element at position (i, j) is the normalized MPC value between the $i^{th}$ and $j^{th}$ channel in the upper triangle of the matrix. Similarly, position (i, j) in the lower triangular matrix corresponds to normalized MSC. The three bands alpha, beta, and gamma correspond to the RGB values of an image matrix. Thus the data input to the classifier is a 62$\times$62 image. The diagonals of the image generated using this method contained no usable information. To add relevant information to the diagonals, the Differential Entropy values provided in the SEED\_EEG dataset \cite{6695876} are introduced as the 62 diagonal elements in each image.

\section{Feature extraction and Methodology}

The main features used in classifying emotions in this work are
\begin{enumerate}
    \item Mean phase coherence (MPC)
    \item Magnitude squared coherence (MSC)
\end{enumerate}

MPC represents the synchronisation between any two channels of the EEG.\cite{MPC,panachakel2021classification} MPC across the $i^{th}$ and $k^{th}$ channel is given by:

\begin{equation}
MPC_{i,k} = \frac{1}{N}\Bigg|\sum_{n=0}^{N-1} e^{-j(\phi_{i}({n}) - \phi_{k}({n}))}\Bigg|
\end{equation}

where $\phi_{i}$(n) and $\phi_{k}$(n) denote the phase of $i^{th}$ and $k^{th}$ channel and N denotes the length of time series data contained in the channel. The phase of each channel is computed using Hilbert Transform.\\
% The MPC is obtained separately for the 3 frequency bands(alpha, beta and gamma) thereby generating 3 matrices for each trial.  

Magnitude squared coherence(MSC) demonstrates the linear relationship between a pair of signals. The MSC between $X_{i*}$ and $X_{k*}$ is given by:
\begin{equation}
MSC_{i,k}({\omega}) = \frac{|S_{i,k}|^{2}}{S_{i,i}({\omega})S_{k,k}({\omega})}
\end{equation}
where \begin{math} S_{i,i}({\omega})\end{math} and\begin{math} S_{k,k}({\omega})
\end{math} denote the power spectral densities and \begin{math} S_{i,k}({\omega})\end{math} is the cross power spectral density.\\
Values of MPC and MSC lies in [0,1].
MPC and MSC for all the combinations of the channels were calculated. MPC was obtained for each of the three bands separately, while MSC was summed and normalized. Since the coherence (i, j) is the same as (j, i), We stacked these features into a matrix of dimension 62$\times$62 such that the upper triangle of the matrix was filled with MPC values and the lower triangle with MSC values. Fig 1. shows the block diagram of the method employed to generate the dataset. The diagonal elements of these matrices contained information about the coherence between the same channels. Hence they carried no information in particular. In \cite{6695876}, the authors concluded that DE is well suited for emotion recognition. Motivated by this, we took advantage of the DE feature provided along with SEED.\\
Let $X$ be a random variable with probability distribution $f$, the DE $h(X)$ is defined as 
% ResNet50 accepts 3 dimensional arrays with each dimension corresponding to one of the EEG frequency bands.

% Thus for each trial three matrices are generated corresponding to alpha, beta and gamma bands and they are stacked into a 3-D matrix of RGB values, thereby generating a pixelated colour image.

% The SEED\_EEG dataset provides the extracted feature Differential entropy (DE) which was first proposed in {\cite{q}} according to which DE is defined as 

\begin{equation}
h(X) = -\int_{-\infty}^{\infty} f(X)\log f(X)dx\\
% &=\frac{1}{2}log(2\pi e\sigma^2)
\end{equation}
We obtained the mean of the DE values for every channel and stacked it into the diagonal of the matrices. 
The matrices were then converted to images such that alpha($\alpha$), beta($\beta$), and gamma($\gamma$) bands correspond to RGB values. The images were then fed to the ResNet-50-based transfer learning model. We performed subject-dependent and subject-independent classifications of emotions.\\
In subject-dependent classification, to avoid overfitting due to training on the limited availability of data, 10-fold cross-validation was employed. We added a global spatial average pooling layer, a fully-connected layer, and a logistic layer to the model as shown in Fig 3.. These layers were randomly initialized and trained on the subject-dependent classification task.\\
In the subject-independent classification process, as shown in Fig 2., we incorporated a 2D convolution layer, a Dense layer, a Flatten layer and finally a Dense layer which generated a probability distribution over the 3 classes.

\section{Experiments}
Two experiments were conducted on the SEED dataset. The first experiment was a subject-dependent classification
%------------#Figure 1--------------
\begin{figure*}[!ht]
\centering
\includegraphics[width=\textwidth]{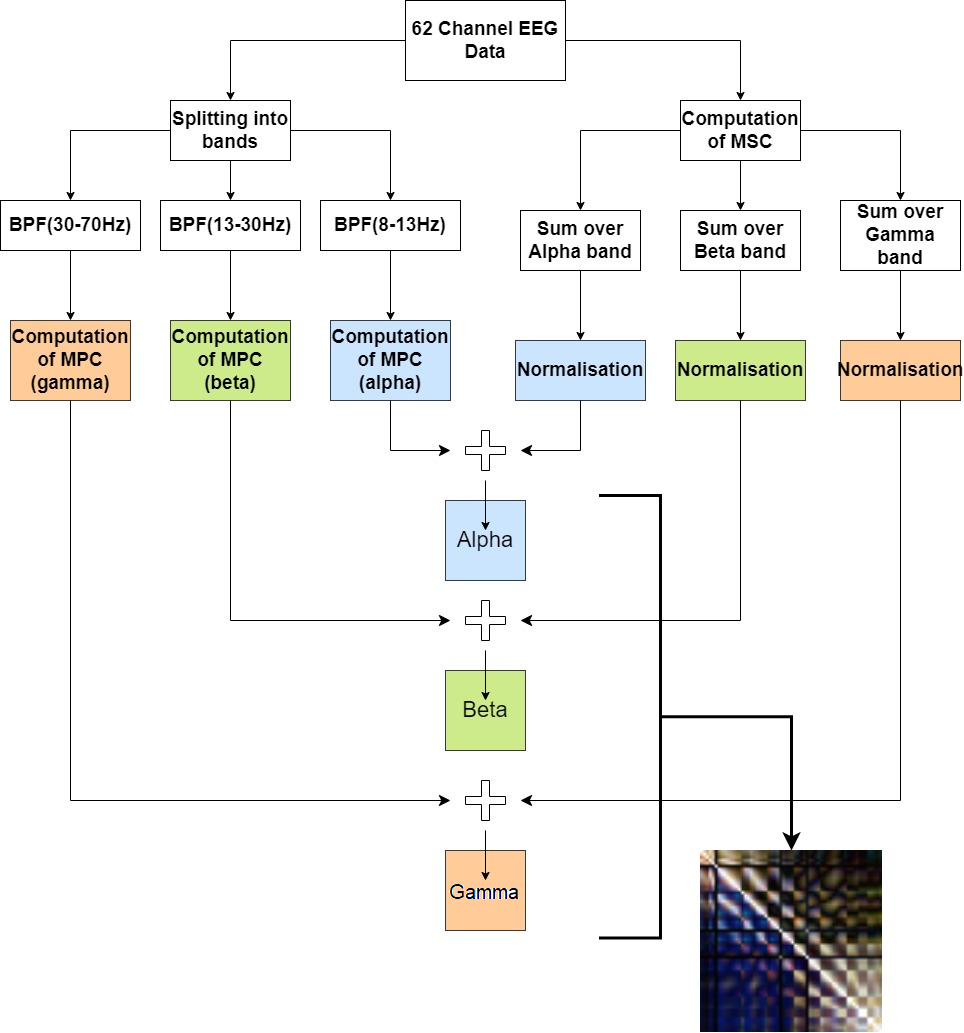}
\caption{Method used for generating the dataset which is fed to the Classifiers.(BPF refers to band pass filter)}
\label{fig:label_for_image}
\end{figure*}

%------------#Figure 2--------------
\begin{figure*}[!ht]
\centering
\includegraphics[width=\textwidth]{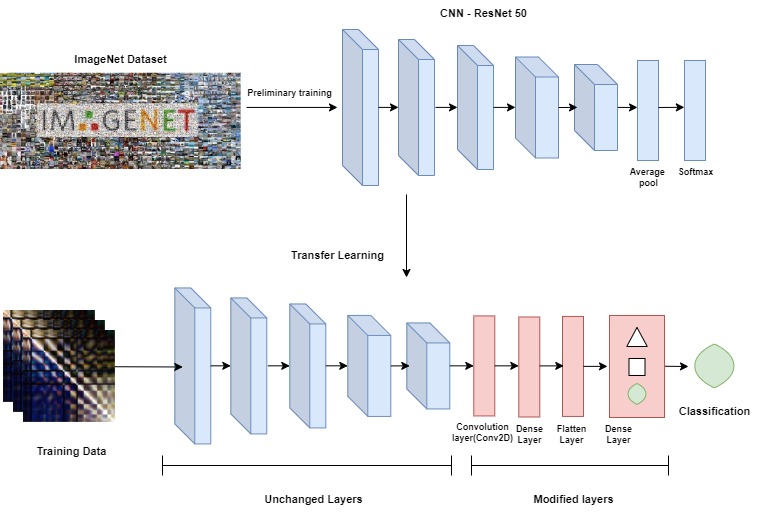}
\caption{Transfer learning approach using ResNet-50 for subject-independent classification}
\label{fig:label_for_image}
\end{figure*}

% \begin{figure}[h]
%         \centering
%         \includegraphics[scale=0.34]{Untitled Diagram.jpg}
%         \caption{Transfer learning approach using ResNet-50 for subject-independent classification}
%     \end{figure}

%#_______________________________________

%------------#Figure 1--------------
% \begin{figure}[!t]
%         \centering
%         \includegraphics[scale =0.8  ]{Block diagram prepro.png}
        
%         \caption{Method used for generating the dataset which is fed to the Classifiers.(BPF refers to band pass filter)}
%     \end{figure}

and the second was a subject-independent classification.\\
1) Subject-dependent experiment:\\
Subject-dependent classification is a method where the classifier is trained and tested on data that is specific to certain subjects or individuals. In this scenario, a 10-fold cross-validation technique was employed to train and evaluate the classifier. The data was divided into 10 equal parts, with 9 parts being used for training and the remaining part being used for testing. This approach allows for a robust evaluation of the classifier's performance, as it is trained and tested on different subsets of the data, simulating the scenario where the model is trained on data from some subjects and tested on data from other subjects.
The evaluation metric used was accuracy. Mean accuracy was obtained by averaging the accuracy scores of all the folds.\\
2) Subject-independent experiment:\\
Subject independent classification refers to a scenario where the classifier is trained and tested on data from different subjects or individuals. In this case, the classifier is not specifically tailored to the characteristics of the subjects for which it is trained and is expected to generalize well to new data from unseen subjects. The experimental setup involved dividing the dataset into a training set and a testing set in a specific way. A leave-one-subject-out (LOSO) cross-validation approach was used, where out of the 15 subjects, one subject was designated as the testing set while the remaining 14 subjects were used as the training set. This process was repeated for each subject so that each subject appears in the testing set exactly once. The mean accuracy obtained from each iteration was calculated, providing an overall measure of the classifier's performance on the dataset.

\section{Results}
We conducted two types of experiments on the SEED dataset: subject-dependent and subject-independent. In the
%#---------Figure 3 ---------------------

\begin{figure*}[!ht]
\centering
\includegraphics[width=\textwidth]{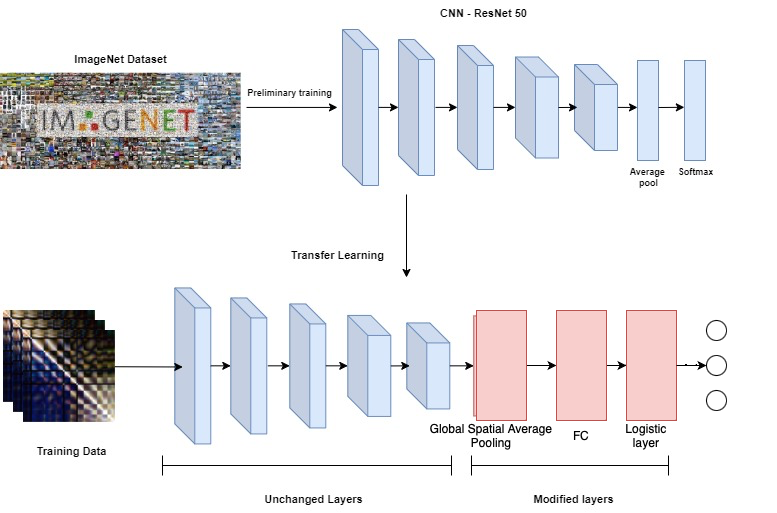}
\caption{Transfer learning approach using ResNet-50 for subject-dependent classification}
\label{fig:label_for_image}
\end{figure*}

% \begin{figure}[h]
%         \centering
%         \includegraphics[scale=0.34]{Untitled Diagram.drawio-4.png}
%         \caption{Transfer learning approach using ResNet-50 for subject-dependent classification}
%     \end{figure}

subject-dependent experiment, the classifier achieved an accuracy of 93.1\% with a standard deviation of 3\%. In the subject-independent experiment, the accuracy obtained was 71.6\% with a standard deviation of 7.6\% as shown in Fig 4. The error bars in the graph shows the standard deviation. The results of these experiments were compared against various previous models, and the comparison is presented in tables 1 and 2. 
\begin{table}[h]
\centering
\caption{Comparison with existing literature for subejct-dependent emotion classification using SEED\_EEG}
\begin{tabular}{|c|c|c|}
\hline
\textbf{Paper} & \textbf{Model} & \textbf{Accuracy} \\ [0.5ex] 
\hline \hline
Liu, Junxiu et al.\cite{Liu2020Sep}   & DNN and SAE & 96.7\%\\
       \hline
      Song, Zheng et al.\cite{8320798} & DGCNN & 90.4\%\\
       \hline
      Asghar, Adeel et al.\cite{9043994} & DNN & 91.3\%\\
       \hline
      Our Model & Transfer learning & 93.1\%\\[1ex]
    \hline
\end{tabular}
\label{tab:table1}
\end{table}

\begin{table}[!h]
\centering
 
 \caption{Comparison with existing literature for subject independent emotion classification using SEED$\_$EEG}
 \begin{tabular}{|c|c|c|} 
 \hline
 \textbf{Paper} & \textbf{model} & \textbf{Accuracy} \\ 
 \hline\hline
 Sunhee Hwang et al.\cite{75.31}  & Multi-task DNN& 75.3\%\\
 \hline
 Cimtay and  Ekmekcioglu.\cite{9195813}   & SVM and DNN & 78.3\%\\
  \hline
 Li, Jingkong et al.\cite{Li2021Jun} & SOGNN & 86.8\%\\
 \hline
 Our Model & Transfer learning & 71.6\% \\[1ex] 
 \hline

\end{tabular}
\vspace{0.2cm}

\end{table}
Cohen's Kappa $\kappa$ is defined as \\
\begin{equation}
\kappa := \frac{p_{cl}-p_{ch}}{100-p_{ch}}
\end{equation}\\
Here, \begin{math}
   p_{cl}
\end{math} is the accuracy of the classifier and Here, \begin{math}
    p_{ch}
\end{math} is the chance level accuracy. The value of \begin{math}
    \kappa
\end{math} lies in the range [-1,1] where values closer to 0 indicate that the performance 
%#------Figure 4 -------------
\begin{figure*}[!ht]
\centering
\includegraphics[width=\textwidth]{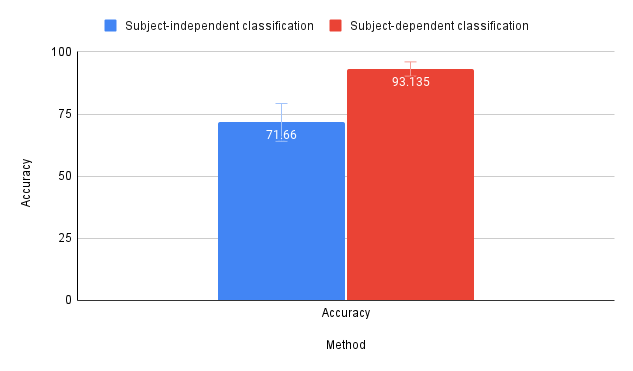}
\caption{Accuracy for subject-dependent and subject-independent classifications}
\label{fig:label_for_image}
\end{figure*}

% \begin{figure}[h]
%         \centering
%         \includegraphics[scale=0.35]{chart-4.png}
%         \caption{Accuracy for subject-dependent and subject-independent classifications}
%     \end{figure}
of the classifier is only as good as a random guess.\\
 The chance accuracy for classifying three classes is 33.33\%.\\
 The $\kappa$ value for the subject-independent classification accuracy is 0.574 and 0.897 for subject-dependent classification.\\
 Our results show that the classifier performed at least twice as well as chance accuracy in both subject-dependent and subject-independent experiments.

\section{conclusion}
% The paper proposes the novel feature extraction technique of combining MPC, MSC and DE for emotion classification which . The best accuracy achieved in Binary classification of emotion (positive, negative) is 93.750$\%$ and is state-of-the-art to the best our knowledge. Further improvements may be made in the accuracy by performing data augmentation.
% The methods described in this paper may be translated to other BCI related fields such as Motor Imagery, Imagined speech classification etc.
In this work, a novel combination of features is proposed for classifying emotions from EEG data. To overcome the limitation of limited data availability, transfer learning using a pre-trained ResNet-50 model is employed. The achieved accuracy in subject-dependent classification is 93.1\%, and in subject-independent classification is 71.6\%. These results are significantly better than the chance accuracy of classifying three classes. The innovation in this work is the use of MPC, MSC and DE as the primary features for EEG emotion classification and compactly arranging them in a three-dimensional array. This three-dimensional array, resembling a colour image in the RGB space is fed to a ResNet-50 model, pretrained on ImageNet dataset, enabling the model to effectively classify emotions.
\section{Future scope}
The future direction of this research includes exploring the potential of data augmentation techniques to enhance the model's performance and generalization ability. By implementing data augmentation, the limitation of limited data availability can be mitigated to some extent. Furthermore, experimenting with different classifiers and a combination of enhanced features can lead to an improvement in classification accuracy. The significance of this research is substantial as it holds the potential to be applied in the field of neuroprosthetics to aid individuals dealing with emotional disorders. In summary, this study presents a novel combination of features for the classification of emotions from EEG data, and the outcomes demonstrate the promise of the proposed method for further research in this area.
\section{Ethics Statement}
Publicly available dataset (SEED\_EEG) was used in this study(https://bcmi.sjtu.edu.cn/home/seed/seed.html). The studies involving human participants were reviewed and approved by Ethics Committee of South China Normal University. The patients/participants provided their written informed consent to participate in this study. Written informed consent was obtained from the individual(s) for the publication of any potentially identifiable images or data included in this article.

\bibliographystyle{unsrt}
\bibliography{bibliography.bib}

\end{document}